\documentclass[aip,reprint]{revtex4-2}
\usepackage{graphicx}
\usepackage{dcolumn}
\usepackage{bm}
\usepackage[dvips]{color}
\usepackage[mathlines]{lineno}
\usepackage{float}
\DeclareUnicodeCharacter{2009}{\,+} 
\begin{document}

\title{Magnetism in four-layered Aurivillius Bi$_5$FeTi$_3$O$_{15}$ at high pressures : A nuclear forward scattering study}

\author{Deepak Prajapat}
\affiliation{UGC-DAE Consortium for Scientific Research, University Campus, Khandwa Road, Indore 452001, India.}

\author{Akash Surampalli}
\affiliation{UGC-DAE Consortium for Scientific Research, University Campus, Khandwa Road, Indore 452001, India.}

\author{Anjali Panchwanee}
\affiliation{Deutsches Elektronen-Synchrotron - A Research Centre of the Helmholtz Association, Hamburg 22607, Germany.}

\author{Carlo Meneghini}
\affiliation{Dipartimento di Scienze, Universita di Roma Tre, I-00146 Roma, Italy}

\author{Ilya Sergeev}
\affiliation{Deutsches Elektronen-Synchrotron - A Research Centre of the Helmholtz Association, Hamburg 22607, Germany.}

\author{Olaf Leubold}
\affiliation{Deutsches Elektronen-Synchrotron - A Research Centre of the Helmholtz Association, Hamburg 22607, Germany.}

\author{Srihari Velaga}
\affiliation{High Pressure and Synchrotron Radiation Physics Division, Bhabha Atomic Research Centre, Mumbai 400085, India.}

\author{Marco Merlini}
\affiliation{Dipartimento di Scienze della Terra "A. Desio", Universita degli Studi di Milano, 20133 Milano, Italy}

\author{Konstantin Glazyrin}
\affiliation{Deutsches Elektronen-Synchrotron - A Research Centre of the Helmholtz Association, Hamburg 22607, Germany.}

\author{Ren{\'e} Steinbr\"{u}gge}
\affiliation{Deutsches Elektronen-Synchrotron - A Research Centre of the Helmholtz Association, Hamburg 22607, Germany.}

\author{Atefeh Jafari}
\affiliation{Deutsches Elektronen-Synchrotron - A Research Centre of the Helmholtz Association, Hamburg 22607, Germany.}

\author{Himashu Kumar Poswal}
\affiliation{High Pressure and Synchrotron Radiation Physics Division, Bhabha Atomic Research Centre, Mumbai 400085, India.}

\author{V. G. Sathe}
\affiliation{UGC-DAE Consortium for Scientific Research, University Campus, Khandwa Road, Indore 452001, India.}

\author{V. Raghavendra Reddy}
\email{varimalla@yahoo.com; vrreddy@csr.res.in}
\affiliation{UGC-DAE Consortium for Scientific Research,  University Campus, Khandwa Road, Indore 452001, India.}

\begin{abstract}
We report the structural and magnetic properties of four-layer Aurivillius compound Bi$_5$FeTi$_3$O$_{15}$ (BFTO) at high hydrostatic pressure conditions.  The high-pressure XRD data does not explicitly show structural phase transitions with hydrostatic pressure, however the observed changes in lattice parameters indicate structural modifications at different pressure values. In the initial pressure region values, the lattice parameters $\textit{a}$- and $\textit{b}$- are nearly equal implying a quasi-tetragonal structure, however as the pressure increases $\textit{a}$- and $\textit{b}$- diverges apart and exhibits complete orthorhombic phase at pressure values of about $\geq$8 GPa. Principal component analysis of high pressure Raman measurements point out an evident change in the local structure at about 5.5 GPa indicating that the evolution of the local structure under applied pressure seems to not follow crystallographic changes (long range order). Nuclear forward scattering (NFS) measurement reveal the development of magnetic ordering in BFTO at 5K with high pressures. A progressive increase in magnetic order is observed with increase in pressure at 5K. Further, NFS measurements carried out at constant pressure (6.4GPa) and different temperatures indicate that the developed magnetism disappears at higher temperatures (20K). It is attempted to explain these observations in terms of the observed structural parameter variation with pressure. 
 
\end{abstract}

\keywords {Bi$_5$FeTi$_3$O$_{15}$, Multiferroic, Nuclear forward scattering, X-ray diffraction, Raman spectroscopy}

\maketitle \section{Introduction} 
Materials which manifest simultaneous magnetic and electric ordering are explored intensively because of their potential applications such as data storage devices, sensing devices etc.\cite{spaldin, zheng2004}. Among the most studied compounds, perovskite BiFeO$_3$ is well explored from the multiferroic/magnetoelectric point of view\cite{wang2003, kothari}. BiFeO$_3$ exhibits very high Neel temperature (T$_N$ $\approx$ 643K) and ferroelectric (FE) Curie Temperature (T$_C$ $\approx$ 1103 K) \cite{fischer, wang2003}. The origin of high temperature ferroelectric (FE) ordering in BiFeO$_3$ is attributed to the presence of 6S\textsuperscript{2} lone pair of electrons\cite{khomskii}. 

Some Aurivillius compounds are also proposed to show similar behavior as they do show FE ordering till very high temperature with low fatigue value and are known as bismuth layered structure ferroelectrics (BSLF) \cite{axiel, axiel1, araujo, park, zuo}.  Aurivillius compounds are a family of layered compounds which have $\textit{n}$ pseudo-perovskite layers of (A$_{n-1}$B$_n$O$_{3n+1}$)$^{2-}$ between the (Bi$_2$O$_2$)$^{2+}$ layers, where A = Na$^+$, Ba$^{2+}$, Ca$^{2+}$, Sr$^{2+}$, Bi$^{3+}$ and B= Fe$^{3+}$, Cr$^{3+}$, Ti$^{4+}$, Nb$^{5+}$, V$^{5+}$, Ta$^{5+}$, W$^{6+}$, Mo$^{6+}$ etc\cite{newnham, axiel, wachsmuth, bai, chen, deepak}. In particular the Aurivillius compounds having magnetic cation (3d metal) are considered as potential candidate for multiferroic behavior. Among them, the four-layer Aurivillius compound BFTO having $\textit{n}$ = 4, A = Bi$^{3+}$, B = Fe$^{3+}$/Ti$^{4+}$ is theoretically reported to be a potential candidate for the multiferroic behavior \cite{axiel, axiel1}. BFTO exhibits FE A2$_1$am(36) crystallographic structure at room temperature which transforms  to paraelectric I4/mmm at very high temperature of the order of 1000 K\cite{li}. The experimental reports suggest that the compound is paramagnetic in nature down to the lowest temperature (2 K) and magnetic properties can be tuned either with pressure or external substitutions such as Nb$^{5+}$, Ta$^{5+}$, W$^{6+}$, Mo$^{6+}$ etc \cite{axiel, axiel1}. 

\begin{figure*}[!]
\centering
\includegraphics[width=\textwidth]{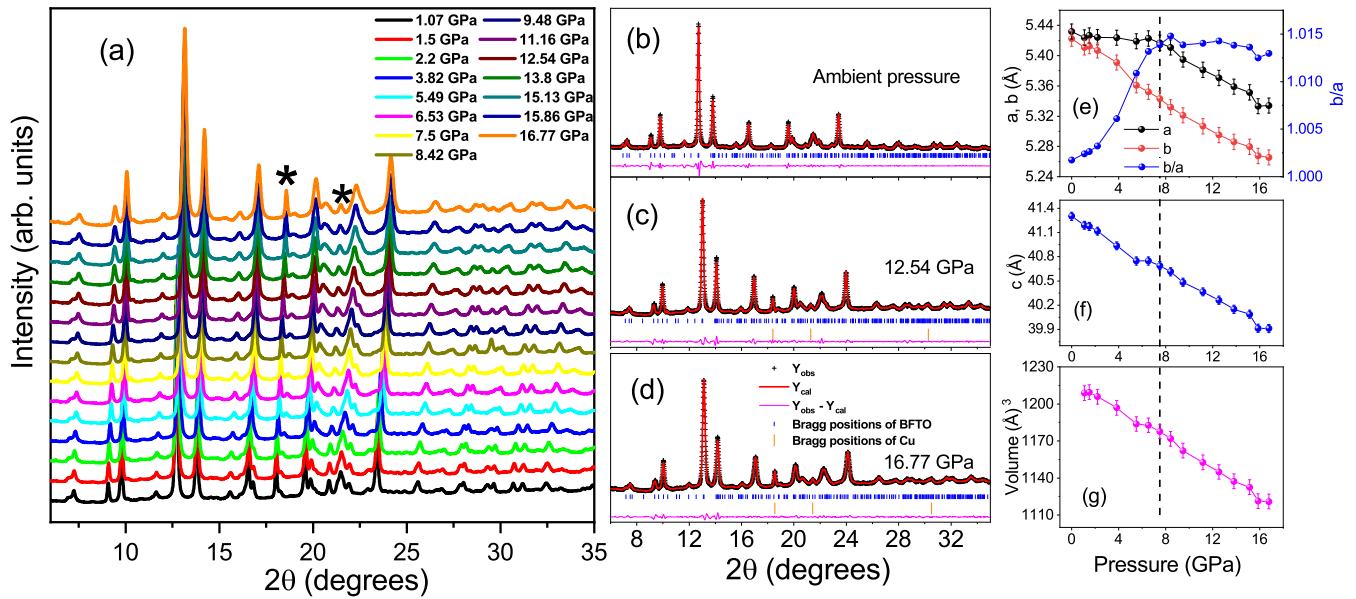}
\caption{(a) Representative pressure dependent x-ray diffraction patterns of polycrystalline Bi\textsubscript{5}FeTi\textsubscript{3}O\textsubscript{15} measured using Indus-2 synchrotron radiation source. Pressure marker (Cu) peaks are shown by * symbol. (b),(c) and (d) Le Bail refined XRD patterns at the selected pressure points. (e), (f) and (g) show the variation of lattice parameters with pressure.}
\label{fig:XRD}
\end{figure*}

The development of magnetic order and hence the multiferroic behavior in BFTO is reported with A/B site substitutions such as La$^{3+}$, Co$^{3+}$, Cr$^{3+}$, Mn$^{3+}$, Ni$^{3+}$, V$^{5+}$ etc\cite{deepak, dias, jartych, mao, zuo, tang, chen1}. However, one of the main problem with A/B site substitutions is the possibility of secondary phase segregation, and consequently stoichiometric modifications, owing the complex phase diagram of Bi-Fe-O \cite{Deepak_HFI}. Therefore, squeezing the lattice, through hydrostatic pressure (or epitaxial strain in thin films) could be the suitable option to explore the possibility of magnetic ordering in BFTO. Because of limited choice of suitable single crystal substrates, the epitaxial strains are not widely explored in BFTO thin films in literature \cite{PLD_NGO}.  Similarly, except the recent work of Ferreira et al., no hydrostatic pressure studies are reported in BFTO \cite{ferreira}. Noticeably, the recent report of Ferreira et al., mainly focused on high pressure structural properties using Raman spectroscopy and x-ray diffraction measurements \cite{ferreira}. While the evolution of magnetic properties of BFTO as a function of pressure remains unexplored. To fill this gap we carried out high pressure investigations in BFTO combining structural (XRD and Raman) and magnetic (M\"{o}ssbauer) probes. In particular Synchrotron based M\"{o}ssbauer measurements (nuclear forward scattering, NFS) are ideal to probe the magnetism at high pressures \cite{NFS_Intro, NFS_HP, NIS_Ilya, NIS_Wenli}. It is observed that magnetic order develops in BFTO at low temperature (5 K) and high hydrostatic pressure. Furthermore the analysis of Raman and XRD reveals a peculiar evolution of long range and local order as a function of the hydrostatic pressure.

\section{Experimental Methods} 

Polycrystalline four-layer Aurivillius compound Bi$_5$FeTi$_3$O$_{15}$ (BFTO) is prepared via solid state reaction method.  High purity (\textgreater 99.9\%) precursors viz., Bi$_2$O$_3$, \textsuperscript{57}Fe\textsubscript{2}O\textsubscript{3} and TiO$_2$ are mixed together in stoichiometric ratio and grinded for 5 hours.  The powder was calcined at 700$^{\circ}$C and then again grinded for 3 hours. The powder was pressed into circular pallets and sintered at 900$^{\circ}$C for 4 hours. High pressure x-ray diffraction (XRD) experiments have been carried out in angle dispersive mode of extreme conditions x-ray diffraction (ECXRD) beamline (BL-11) at Indus-2, INDUS 2 synchrotron RRCAT Indore, India using a Mao–Bell type diamond anvil cell (DAC).  The powder sample, along with copper (Cu) metal powder (as a Pressure Calibrant) is loaded in a gasket hole (with a diameter of $\approx$ 250 $\mu$m) drilled in a pre-indented tungsten gasket. The hydrostatic environment is maintained using a methanol+ethanol (4:1) pressure transmitting medium. High-pressure XRD experiments have been performed up to 16.8GPa using an x ray of wavelength 0.6523 \AA. For refining the wavelength and sample to the detector distance, a diffraction pattern of the NIST (the National Institute of Standards and Technology, USA) standard, LaB\textsubscript{6}, is used. The diffraction data were collected using a MAR345 image plate area detector. Obtained images are converted into one-dimensional diffraction profiles using FIT2D\cite{FIT2D} software. The Raman spectroscopy measurement was performed using Jobin Yvon Horiba LABRAM spectrometer equipped with 632.7 nm laser source to excite Raman signal and by placing the sample in DAC up to 14.31 GPa pressure values. Stainless steel was used as the gasket and a hole of around 150$\mu$m was drilled for sample mounting.  Ruby powder was mixed with the sample to determine the pressure values during the Raman measurement. \textsuperscript{57}Fe M\"{o}ssbauer measurements were carried out in transmission mode using a standard PC-based M\"{o}ssbauer spectrometer equipped with a WissEl velocity drive in constant acceleration mode. The spectrometer was calibrated with natural iron at room temperature. High pressure and low temperature nuclear forward scattering (NFS) measurements were performed at beamline P-01, PETRA III, DESY, Hamburg. The compact DACs are made from CuBe alloy and were installed into cryomagnetic system allowing for measurements down to about 3 K \cite{NIS_Ilya}. In order to assure that we measure from the same region of the sample, the high pressure cells were aligned to the beam in a way that we always hit the same part of the sample. The photon beam was focused to about 8 $\mu$m x 10 $\mu$m (vert. x hor.).

\section{Results and Discussion}

High-pressure XRD measurements are shown in Figure~\ref{fig:XRD}. The diffraction patterns are qualitatively similar to that of Ferreira et al., \cite{ferreira} but the present data covers larger angular range providing better precision on the refined structural parameters. The data is analyzed with Le Bail method using FullProf software \cite{fullprof}. The ambient pressure XRD pattern is fitted considering A2$_1$am(36) space-group and the observed lattice parameters at ambient pressure are $\textit{a}$=5.43(2)\AA, $\textit{b}$=5.42(2)\AA and $\textit{c}$=41.30(1)\AA, which excellently match with the earlier reports\cite{Deepak_JPCM}. Raising the pressure, the XRD patterns are also observed to fit with the same space group (A2$_1$am) indicating no structural transitions (Figure~\ref{fig:XRD} c,d) till the highest pressure values in the present work. The Le Bail refinement method, not considering the unit cell structure, provides higher precision on the lattice parameters. Rietveld refinement\cite{fullprof}  have been carried out at selected pressure points. Owing the number of in-equivalent atomic positions in the unit cell and consequently the large number of free parameters in the refinement, the unit cell structure (atomic position) is affected by some larger uncertainty and would require deeper investigation.  However the A2$_1$am(36) space group is confirmed with the same lattice parameters as from the Le Bail Analysis. The relevant structural parameters, Rietveld fitted XRD patterns for representative pressure values are shown in supplementary data \cite{Supple} and the goodness of fit parameter indicate that the A2$_1$am(36) space-group is a correct choice in fitting the high pressure XRD data, as shown in Figure~\ref{fig:XRD}. 

The obtained lattice parameters decrease as pressure increases as shown in Figure~\ref{fig:XRD}. The effect of pressure is different on the three axes. In particular the rate of their decrease with pressure (d($\textit{a,b,c}$)/dP) is different.  Quantitatively the average slope of lattice parameter change as a function of pressure are respectively d$\textit{a}$/dP = -0.006(1),  d$\textit{b}$/dP = -0.009(1) and  d$\textit{c}$/dP = -0.082(1), evidencing an order of magnitude more sensitivity of $\textit{c}$ parameter with pressure. In addition to this observation, one may also note that in the initial pressure region values, the lattice parameters $\textit{a}$- and $\textit{b}$- are nearly equal and shows a quasi-tetragonal structure. This observation is similar to that of Ferreira et al., \cite{ferreira} wherein for pressures $\leq$ 3.2 GPa the $\textit{a}$- and $\textit{b}$- parameters are close to each other. However, for pressures between about 3.2-7.5 GPa, Ferreira et al., observed almost same values of $\textit{a}$- and $\textit{b}$- and hence a transformation to tetragonal structure is proposed. But our present data depict a different scenario, in fact we noticed that $\textit{a}$- and $\textit{b}$- diverges apart with raising the pressure and behave parallel as a function of pressure above 8 GPa, exhibiting a complete orthorhombic phase. This trend is even more clear looking at the $\textit{b/a}$ ratio as shown in Figure~\ref{fig:XRD}(e).  

Therefore, even though our XRD data does not explicitly show structural phase transitions with hydrostatic pressure the trend in lattice parameter variation indicate structural modifications with hydrostatic pressure across different regions. To further understand such changes observed in XRD, the data is analyzed with principal component analysis (PCA) and the results are discussed in the later part.  In addition to XRD, Raman spectroscopy measurements were also done at high pressures as discussed below. It may be noted that the XRD and Raman data provide widely complementary details about the sample structure, XRD probing the crystallographic structure of the samples (long range order), while the Raman spectroscopy, probing the vibrational and deformation modes of the atomic units, is sensitive to the local atomic order in the sample.

\begin{figure}[t]
\centering
\includegraphics[height = 12 cm, width=8 cm, keepaspectratio]{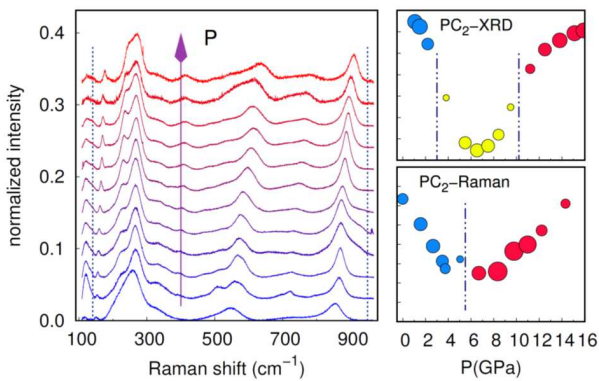}
\caption{left panel: Normalized Raman spectra as a function of pressure (ambient to 14.31 GPa, vertically shifted for sake of clarity), dotted lines specify the region of PCA analysis. right panels: PC$_2$ for XRD (top) and Raman (bottom) as a function of pressure. The colors of the symbols specify the cluster assignment while the symbol size is proportional to the silhouette of the point, quantifying the confidence in the assignment of the point to its cluster. Vertical lines are guide to the eye individuating the critical pressure points.}
\label{fig:PCA1}
\end{figure}

Pressure dependent Raman spectroscopy measurement have been performed in the range 90 cm$^{-1}$ to 960 cm$^{-1}$ and the data is shown in Figure~\ref{fig:PCA1}. The Raman modes are observed to shift towards higher wave-number at high pressure conditions as expected. Raman spectra are fitted considering Lorentzian line shape and it is observed that all the modes broaden with pressure. The variation of Raman mode position with pressure is shown in supplementary data \cite{Supple}.  No visual anomalies were observed in the variation of Raman modes position with pressure which could be due to the relatively large uncertainty in its estimation due to the overlapping of various modes and, partial overlap of the peaks and some correlation among the best fit parameters. In order to have better sensitivity to the pressure effects in the Raman spectra, we applied the principal component analysis (PCA). In our recent works, this approach has been used for the Raman spectroscopy data and demonstrated its sensitivity to the finest structural modifications in BFTO and compounds such as relaxor ferroelectrics across dielectric maxima \cite{Akash_PRB, Deepak_JPCM}. 

The PCA \cite{PCA} being a valuable descriptive tool suitable for the analysis of ample data sets that can be adapted to different contexts, allows to extract significant semi-quantitative information from the variability within a dataset. The principal components (PCs) are the eigenvectors of the covariance matrix of the entire dataset (XRD patterns or Raman spectra in our case). The PCs are ordered so that the formers take into account the largest fraction of the data variance, likely taking into account for real changes within the dataset. On the contrary the effects of statistical noise are likely relegated to the later PCs. Since first few (3-4) PCs often contain most of the relevant information, the PCA simplifies the interpretation of a complex datasets allowing to intuitively visualize the modifications induced by an independent variable, as the applied pressure in the present work. 

Here we applied the PCA to the XRD and Raman spectra independently using Orange software \cite{Orange}. The XRD/Raman data have been imported in the 2$\theta$ range of 6-34$^{\circ}$ and 140-950 cm$^{-1}$ range respectively. The background was subtracted (rubber band model) and the data were normalized (vector normalization). In both the XRD and Raman datasets, it is observed that the first 3 PCs account for more than 95\% of the total variance (please see Figure S3 in supplementary data\cite{Supple}). The XRD/Raman data were partitioned using the K-means \cite{KMeans} algorithm and the Silhouette analysis \cite{Silouette} implemented in the Orange program. The Silhouette analysis indicated three statistically significant clusters for the XRD dataset and only two statistically significant clusters for the Raman dataset as shown in supplementary data \cite{Supple}. 

\begin{figure}[t]
\centering
\includegraphics[width=8 cm]{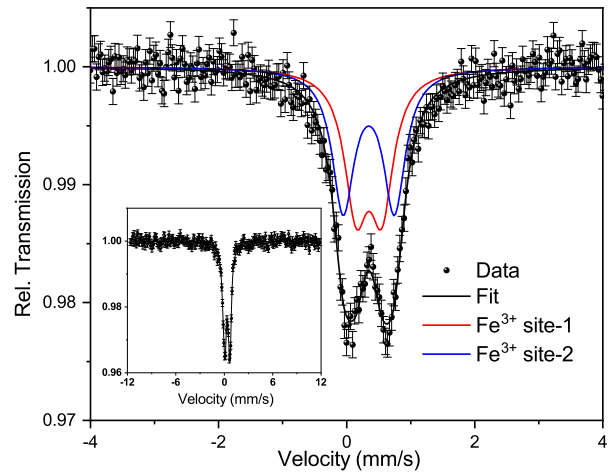}
\caption{Room temperature $^{57}$Fe M\"{o}ssbauer spectra of polycrystalline Bi\textsubscript{5}FeTi\textsubscript{3}O\textsubscript{15}. Inset show the RT data of the same sample measured covering higher velocity range. Symbols represent the experimental data and the solid line is the best fit to the data.}
\label{fig:Moss}
\end{figure}

\begin{figure}[b]
\centering
\includegraphics[width=8 cm]{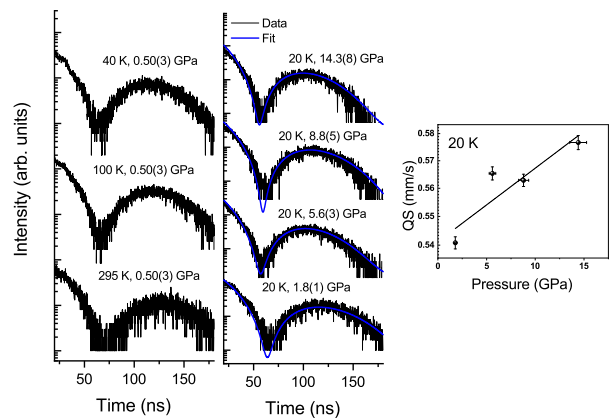}
\caption{Representative NFS spectra of Bi\textsubscript{5}FeTi\textsubscript{3}O\textsubscript{15} measured at the indicated pressure and temperatures. The data depicts the presence of paramagnetic ordering. The data of 20 K is fitted considering paramagnetic doublet and the obtained variation of quadrupole splitting (QS) with pressure is shown in right side frame.}
\label{fig:NFS_Para}
\end{figure}

In Figure~\ref{fig:PCA1} the normalized Raman spectra are shown together with the PC$_2$ for the XRD and the Raman datasets. The analysis of XRD datasets individuates three regions as a function of pressure with a first change around 3 GPa and a second one around 10 GPa in agreement with previous work by Ferreira et al \cite{ferreira}. However, the proposed tetragonal structure at intermediate pressures cannot be confirmed in the present work. The analysis of the Raman dataset individuates only two regions  with an evident change around 5.5 GPa. The PC$_i$ vs PC$_j$ behavior (Figure S3, supplementary) demonstrate the correlation between the principal components and qualitatively motivate the clustering. 

To summarize, both the Raman and XRD data demonstrate a broad smooth structural deformation occurring in the 3-10 GPa region. However the PCA suggests some difference between the evolution of long range order (XRD patterns) or local structure (Raman spectra) as a function of pressure. This finding suggests the evolution of the local structure under applied pressure seems to not follow crystallographic changes (long range order), which needs to be explored with some more measurements which can probe local length scales such as x-ray absorption spectroscopy. 

\begin{figure*}[htbp]
\centering
\includegraphics[height=10cm, width=\textwidth, keepaspectratio]{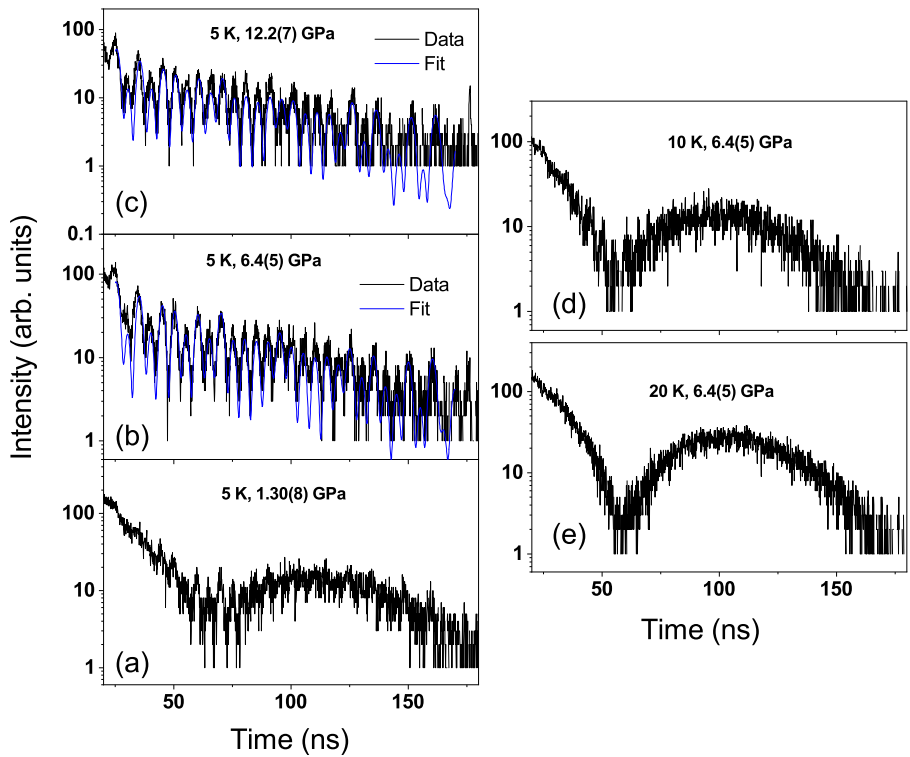}
\caption{NFS spectra of Bi\textsubscript{5}FeTi\textsubscript{3}O\textsubscript{15} measured at the indicated temperature and pressure values. The occurrence of quantum beat patterns clearly reveal the development of magnetic ordering in BFTO at 5K with increasing pressure as shown in (a), (b) and (c). At a constant pressure (6.4 GPa) the magnetic ordering is observed to wipe-out with increasing temperature as shown in (b), (d) and (e).}
\label{fig:NFS_5K}
\end{figure*}

Figure~\ref{fig:Moss} shows the \textsuperscript{57}Fe M\"{o}ssbauer spectra of the BFTO sample measured at room temperature and the data shows paramagnetic doublet. The data is fitted considering two sites corresponding to B1- and B2-sites of BFTO as usually done in the literature \cite{Deepak_JPCM}. It may be noted that both these sites correspond to Fe\textsuperscript{3+} in octahedral coordination and the site with relatively less quadrupole shift value is assigned to the B1-site and the site with relatively higher value of QS is assigned to B2-site in literature \cite{Deepak_JPCM}. Further, in order to check for the presence of iron oxide based magnetic secondary phases in the sample the RT measurements are carried out covering high velocity range. It may be noted that one would expect a magnetic sextet at room temperature if secondary phases such as BiFeO\textsubscript{3} (T\textsubscript{N} $\approx$ 643 K), $\alpha$-Fe\textsubscript{2}O\textsubscript{3} (T\textsubscript{N} $\approx$ 950 K) etc., are present,  but the observed paramagnetic doublet even at higher velocity scale as shown in the inset of Figure~\ref{fig:Moss} conclusively rules out the presence of such phases. Similarly the observed center shift values indicate the presence of only Fe\textsuperscript{3+} in the studied samples and therefore rule out the presence of Fe\textsuperscript{2+}TiO\textsubscript{3} (T\textsubscript{N} $\approx$ 55 K) based phases in the samples \cite{FTO}. The conventional energy-domain \textsuperscript{57}Fe M\"{o}ssbauer measurements are time consuming while the very small amount of sample in high pressure experiments practically limits the signal. Therefore, \textsuperscript{57}Fe M\"{o}ssbauer measurements as a function of pressure at low temperatures have been carried out with synchrotron radiation exploiting the nuclear forward scattering (NFS).

The NFS measurements are carried out at different temperatures with different pressure values. The representative NFS data is shown in Figure~\ref{fig:NFS_Para} and Figure~\ref{fig:NFS_5K}. The NFS data measured from ambient to high pressures show paramagnetic nature till 20 K. The NFS data at 20K as a function of pressure are fitted using CONUSS program \cite{CONUSS} considering only one site as a simple model and the obtained variation of quadrupole splitting (QS) value with pressure till 14.3 GPa is also shown in Figure~\ref{fig:NFS_Para}. The increasing value of QS with pressure is related to the lattice modifications in BFTO. It may be noted that the there are two main contributions to QS value viz., the lattice contribution (QS\textsubscript{lat}) from the crystal field produced by the surrounding ions and the electronic contribution (QS\textsubscript{el}) from the non-spherical charge distribution of its own electron shell \cite{Moss_Book}. Because of the presence of spherically symmetric high-spin Fe\textsuperscript{3+} state in BFTO, QS\textsubscript{lat} can be considered as the main contribution to the measured QS value. 

At the lowest temperature (5 K) the evolution of the NFS signal reveals an interesting behavior as shown in Figure~\ref{fig:NFS_5K}. Raising the pressure, the NFS data progressively develop high frequency oscillations indicative of a magnetic ordering. The spectra at 5 K and lowest pressure ($\approx$1.3 GPa) reveal the convolution of paramagnetic and magnetic features. Raising the hydrostatic pressure to about 6.4 and 12.2 GPa well defined quantum beat pattern indicating the presence of magnetic ordering is observed. Further, to check the variation of magnetic ordering temperature with pressure, the measurements are carried out at a given pressure (6.4 GPa) with increasing temperature. It is observed that while at 10 K and 6.4 GPa a weak magnetic component still persists, the NFS spectrum converts back to complete paramagnetic state at 20 K \& 6.4 GPa as shown in Figure~\ref{fig:NFS_5K}. The lowest temperature high pressure NFS data are quantitatively analyzed with CONUSS\cite{CONUSS} program. However the quantitative analysis of the spectra resulting from a  combination of magnetic and para- magnetic features are excluded because the many free parameters in the fitting gave poorly reliable values. Whereas, the magnetic data is fitted reliably considering two equal area anti-parallel magnetic sites with almost similar hyperfine field values of about 53.8 T fully aligned in a plane perpendicular to the beam and a quadrupole interaction of about 0.45 mm/s perpendicular to the magnetic hyperfine fields. With this model an acceptable fit to the NFS data is obtained as shown in Figure~\ref{fig:NFS_5K}. The two equal area anti-parallel sites fitting suggest the presence of anti-ferromagnetic ordering, however NFS measurements with external magnetic field are needed to confirm the nature of magnetic ordering.  

Development of magnetism with pressure is usually explained in terms of squeezing inter-atomic distances between magnetic atoms favoring the magnetic exchange interactions in compounds which are otherwise non-magnetic. For example, decrease in Fe-Fe distance is believed to be responsible for the development of magnetism with pressure in compounds such as SrFeO\textsubscript{3}. Pressure induced ferromagnetism is reported in SrFeO\textsubscript{3} with high pressure \textsuperscript{57}Fe M\"{o}ssbauer spectroscopy measurements and it is reported that the Neel temperature increases at a rate of about 9 K/GPa and a clear magnetic component developed at room temperature in SrFeO\textsubscript{3} for about 53 GPa \cite{SFO_HP}. A similar mechanism could occur in our BFTO system  i.e., the decrease in inter-atomic distance between Fe might be happening in the present case with high pressure. In this work the Fe(Ti)-Fe(Ti) distance decreases with pressure (Figure S1, Supplementary). However, the relatively low concentration of magnetic cations (only 25\% of Fe\textsuperscript{3+}) and short range of magnetic super-exchange interaction in BFTO as compared to SrFeO\textsubscript{3} might be responsible for stabilizing the magnetic order at the lowest temperature (5 K) and the magnetic order loss as soon as you go up a few degrees K (10 K, Figure~\ref{fig:NFS_5K}).  It is significant that the lattice parameter $\textit{c}$ is much more sensitive to the applied pressure than the axes $\textit{a}$ and $\textit{b}$, this might be effective in varying the interlayer (J\textsubscript{INTER}) coupling in BFTO \cite{axiel1}. However, the strength of J\textsubscript{INTER} is orders of magnitude less as compared to other magnetic coupling mechanisms in BFTO.  The three types of coupling mechanisms considered between magnetic ions in BFTO are nearest-neighbor (J\textsubscript{NN} = 45 meV), next-nearest-neighbor (J\textsubscript{NNN} = 1.35 meV) and interlayer (J\textsubscript{INTER} = 0.45 meV) \cite{axiel1}. As a result it is argued that the most promising route to obtain magnetic long-range order at or above room temperature in BFTO is to increase the concentration of magnetic ions as compared to enhancing the J\textsubscript{NNN} and J\textsubscript{INTER} via pressure etc \cite{axiel1}. Application of very high pressures ($\geq$ 20 GPa or so) might help in achieving room temperature magnetism in BFTO, however the stability of the layered structure at such high pressure also needs to be investigated as the present results suggest that the lattice parameter $\textit{c}$ is more sensitive to pressure.

\section{Conclusions}
In conclusion, detailed analysis of structural data (x-ray diffraction and Raman) with high hydrostatic pressure conditions reveal no structural phase transitions in Bi$_5$FeTi$_3$O$_{15}$ (BFTO) till the studied pressure values ($\approx$ 16 GPa). However the observed changes in lattice parameters and Raman data indicate structural modifications at different pressure values. The main outcome of the present work is the evidence for the development of magnetism in BFTO with high pressure, which is otherwise expected to be non-magnetic. However, the ordering temperature is found to be very low (5 K). Measurements with further high pressures might enable the stabilization of magnetism at higher temperatures in BFTO and hence opening up the possibility the multiferroic behavior in four layer Aurivillius compounds.

\section{Acknowledgments}
VRR would like to acknowledge the support by the Department of Science and Technology (Government of India) provided within the framework of the India@DESY collaboration for the NFS measurements. Authors thank Er. Ajay Rathore and Mr. Binoy Krishna De for help during Raman spectroscopy measurement. DP thank Dr. Ganesh Bera and Er. Anil Gome for help and discussions.

\section{References}

\bibliography{References}

\end{document}


\title{Supplemental material to Magnetism in four-layered Aurivillius Bi$_5$FeTi$_3$O$_{15}$ at high pressures : A nuclear forward scattering study}

\author{Deepak Prajapat}
\affiliation{UGC-DAE Consortium for Scientific Research, University Campus, Khandwa Road, Indore 452001, India.}

\author{Akash Surampalli}
\affiliation{UGC-DAE Consortium for Scientific Research, University Campus, Khandwa Road, Indore 452001, India.}

\author{Anjali Panchwanee}
\affiliation{Deutsches Elektronen-Synchrotron - A Research Centre of the Helmholtz Association, Hamburg 22607, Germany.}

\author{Carlo Meneghini}
\affiliation{Dipartimento di Scienze, Universita di Roma Tre, I-00146 Roma, Italy}

\author{Ilya Sergeev}
\affiliation{Deutsches Elektronen-Synchrotron - A Research Centre of the Helmholtz Association, Hamburg 22607, Germany.}

\author{Olaf Leubold}
\affiliation{Deutsches Elektronen-Synchrotron - A Research Centre of the Helmholtz Association, Hamburg 22607, Germany.}

\author{Srihari Velaga}
\affiliation{High Pressure and Synchrotron Radiation Physics Division, Bhabha Atomic Research Centre, Mumbai 400085, India.}

\author{Marco Merlini}
\affiliation{Dipartimento di Scienze della Terra "A. Desio", Universita degli Studi di Milano, 20133 Milano, Italy}

\author{Konstantin Glazyrin}
\affiliation{Deutsches Elektronen-Synchrotron - A Research Centre of the Helmholtz Association, Hamburg 22607, Germany.}

\author{Ren{\'e} Steinbr\"{u}gge}
\affiliation{Deutsches Elektronen-Synchrotron - A Research Centre of the Helmholtz Association, Hamburg 22607, Germany.}

\author{Atefeh Jafari}
\affiliation{Deutsches Elektronen-Synchrotron - A Research Centre of the Helmholtz Association, Hamburg 22607, Germany.}

\author{Himashu Kumar Poswal}
\affiliation{High Pressure and Synchrotron Radiation Physics Division, Bhabha Atomic Research Centre, Mumbai 400085, India.}

\author{V. G. Sathe}
\affiliation{UGC-DAE Consortium for Scientific Research, University Campus, Khandwa Road, Indore 452001, India.}

\author{V. Raghavendra Reddy}
\email{varimalla@yahoo.com; vrreddy@csr.res.in}
\affiliation{UGC-DAE Consortium for Scientific Research,  University Campus, Khandwa Road, Indore 452001, India.}

\maketitle  

High pressure x-ray diffraction (XRD) data of polycrystalline Bi\textsubscript{5}FeTi\textsubscript{3}O\textsubscript{15} (BFTO) is analyzed with profile matching (Le Bail) as shown in Figure-1 of the main text. However, for few representative pressure values the XRD patterns are refined with Rietveld refinement considering A2$_1$am(36) space-group and the data is shown in Figure.~\ref{fig:XRD}. The relevant parameters such as Wyckoff positions, occupancy etc., are shown in Table. ST1. One can see that high pressure XRD data is also fitting with the same A2$_1$am(36) space-group indicating that there is no structural phase transition in BFTO with hydrostatic pressure ($\leq$ 16.8 GPa). Further, it may be noted that the obtained lattice parameters are observed to be same as that of profile matching results. From the refined XRD data, the variation of Fe(Ti)-Fe(Ti) inter-atomic distance is obtained using the VESTA program\cite{VESTA} and the results are also shown in Figure. S1 along with the schematic of the structure of BFTO. 

\renewcommand{\thefigure}{S1}
\begin{figure}[htbp]
  \centering
  \begin{minipage}[b]{0.32\textwidth}
    \includegraphics[height=6cm, width=\textwidth, keepaspectratio]{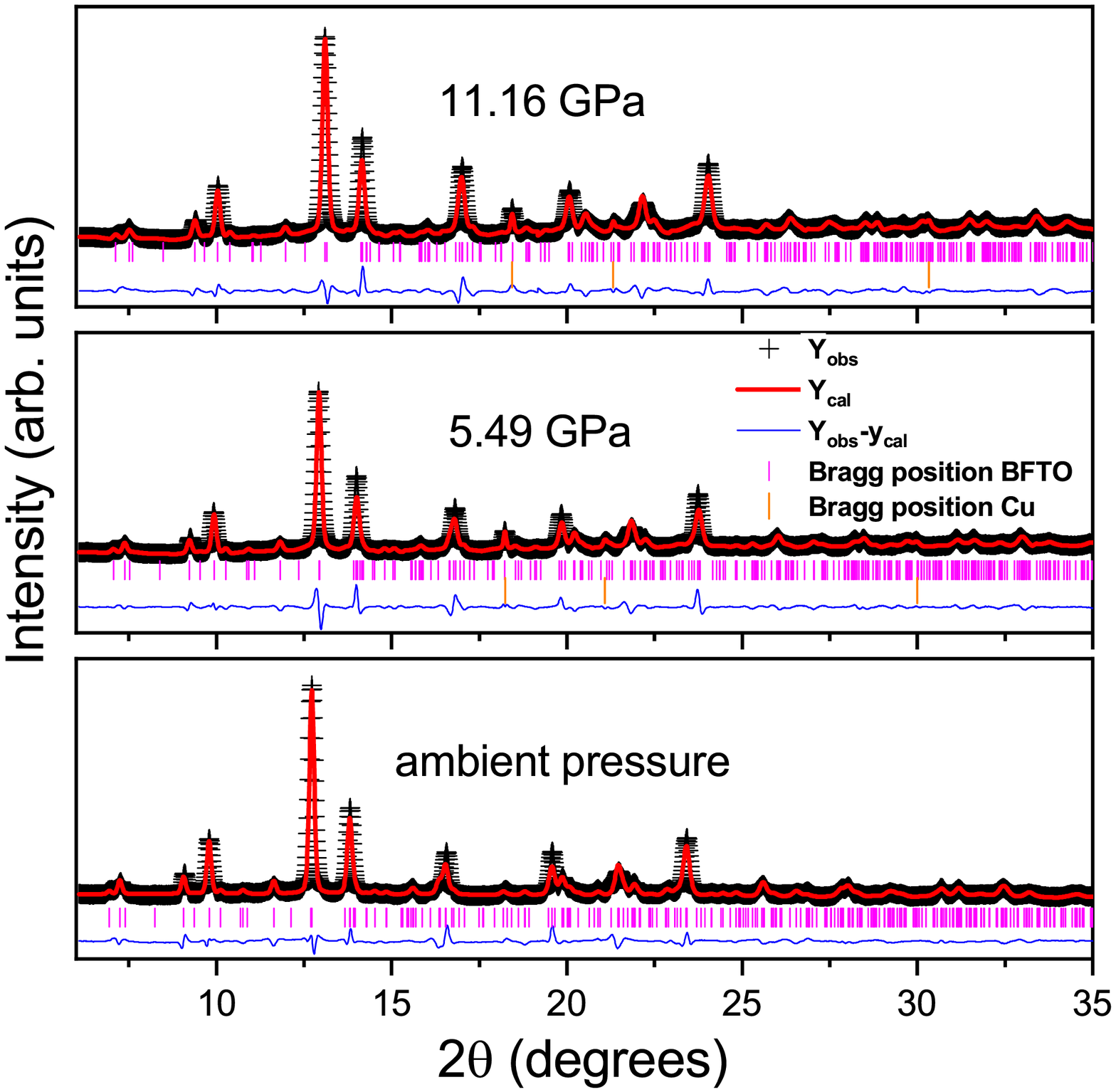}
  \end{minipage}
  \hfill
  \begin{minipage}[b]{0.32\textwidth}
    \includegraphics[height=6cm,width=\textwidth, keepaspectratio]{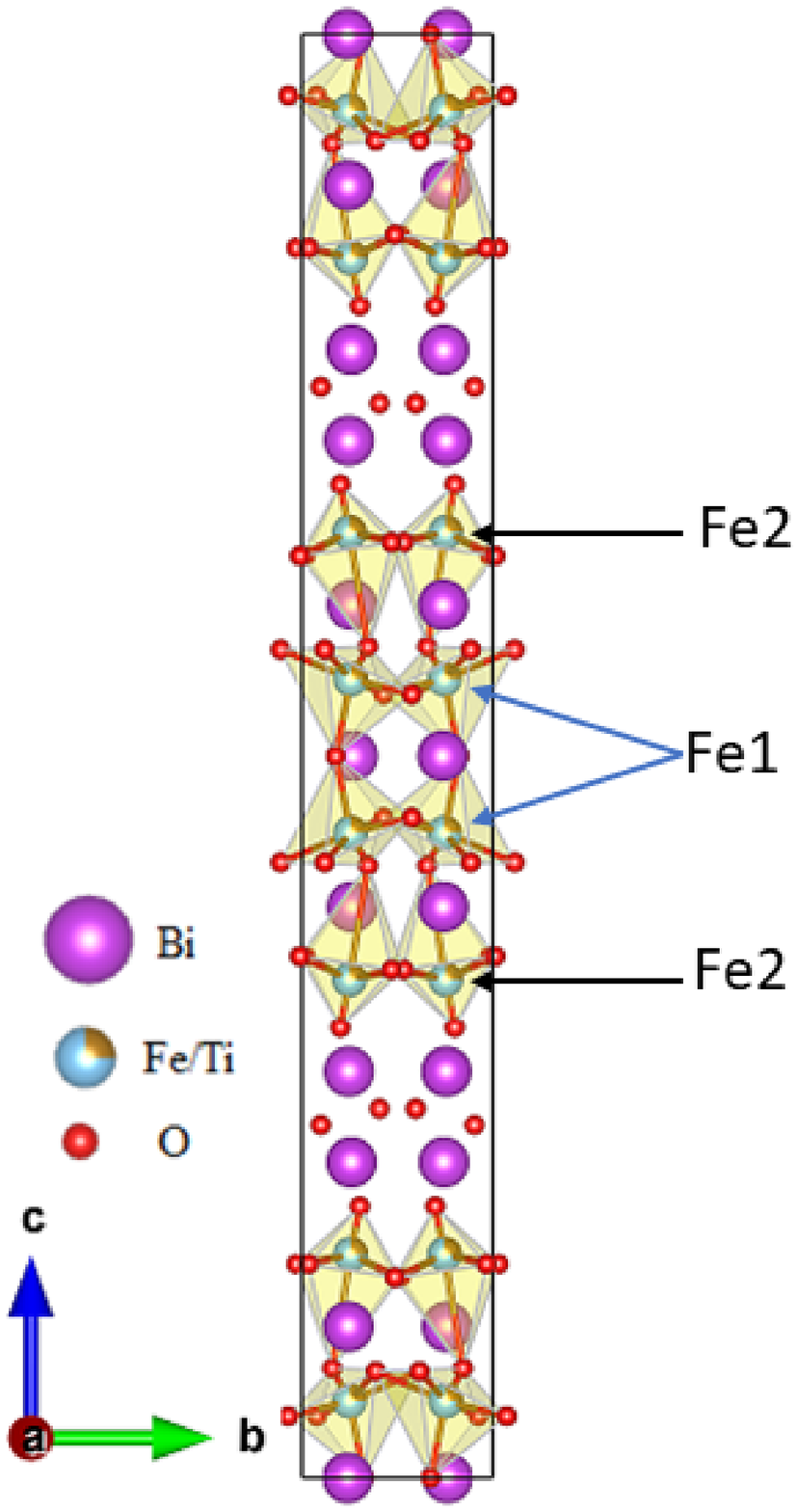}
  \end{minipage}
	\hfill
	\begin{minipage}[b]{0.32\textwidth}
    \includegraphics[height=6cm, width=\textwidth, keepaspectratio]{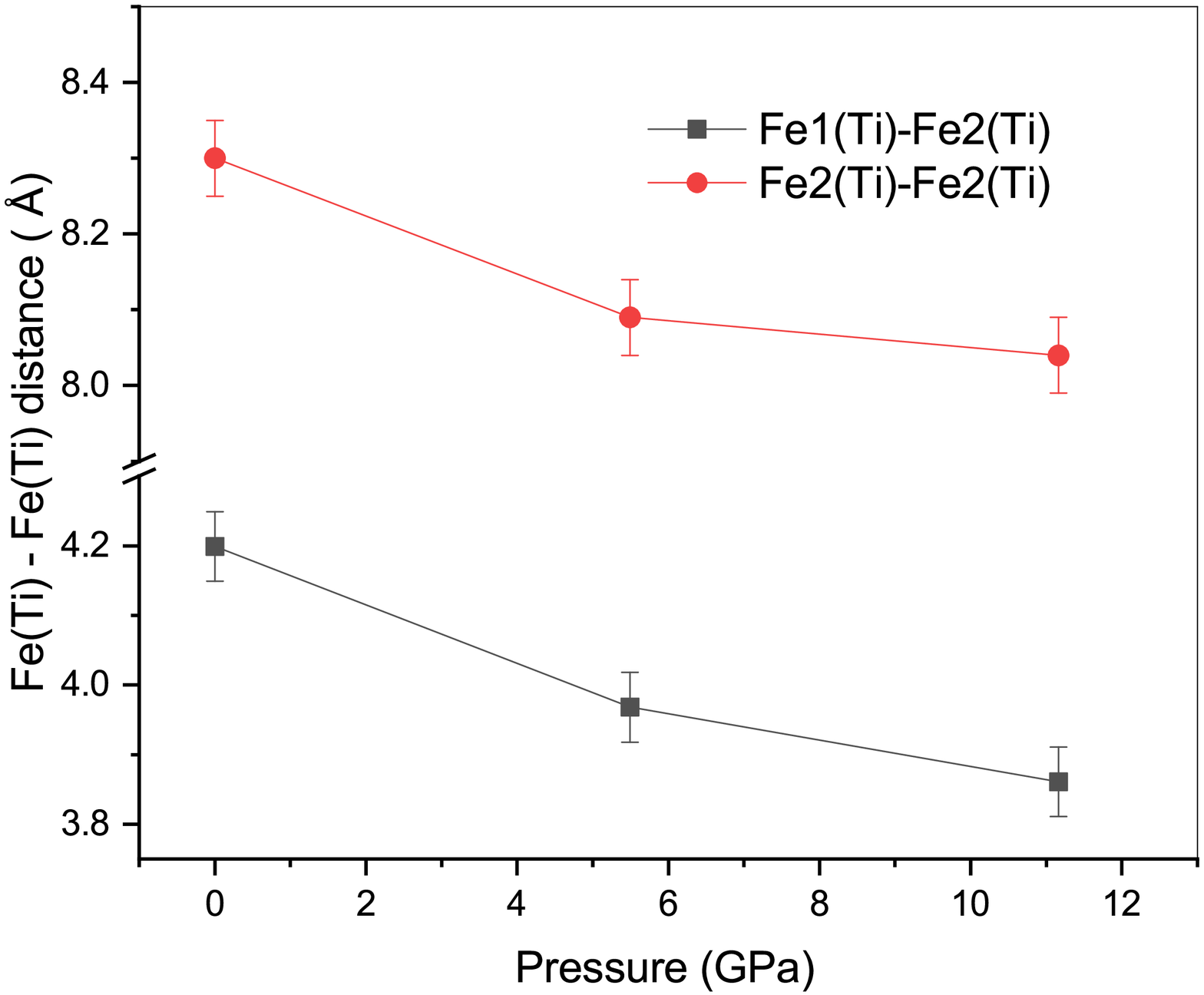}
  \end{minipage}
  \hfill
	\caption{(Left) Representative Rietveld refined XRD patterns of polycrystalline Bi\textsubscript{5}FeTi\textsubscript{3}O\textsubscript{15} (BFTO) at different pressure values (Middle) Structure of BFTO highlighting different Fe atoms (Right) Variation of Fe-Fe inter-atomic distance with pressure obtained from the Rietveld refinement.}
	\label{fig:XRD}
\end{figure}

\renewcommand{\thetable}{ST1}
\begin{table*}[htbp]
\caption{\label{arttype} Structural parameters obtained from XRD for Bi\textsubscript{5}FeTi\textsubscript{3}O\textsubscript{15} at 0.0 GPa, 5.49 GPa and 11.16 GPa, refined with space group A2$_1$am(36) [P1, P2 and P3 are the Wyckoff positions at 0.0 GPa, 5.49 GPa and 11.16 GPa pressure respectively]. }
\begin{ruledtabular}
\begin{tabular}{ccccccccccc}
Atom  & & x &  & & y &  & & z &   & occupancy   \\	
      \cline{2-4}  \cline{5-7}   \cline{8-10}
      &P1 &P2& P3    &P1 &P2 &P3     &P1 &P2 &P3     &              \\
\hline
Bi(1)		& 0.25 &0.25 & 0.25             &0.7628 & 0.7714(6) &0.7643(7)     &0.0 &0.0&0.0 	      &0.50 \\
Bi(2)	  &0.2672(3) &0.2530(9)  &0.2640(3)        &0.7529(3) &0.7310(6) &0.7383(6)	 &0.2185(1) &0.2174(1) &0.2181 	&1.0 \\
Bi(3)		&0.2474(7) & 0.2454(2) &0.2622(5)          &0.7596(1) &0.7674(9) &0.7683(3)  &0.1046(7) &0.1044(6) & 0.1050(7)  	  &1.0  \\
Ti(1)		&0.2605(4) &0.2633(2)  &0.2608(3)           &0.2469(7) &0.2395(1) &0.2142(4) 	   & 0.0532(1) &0.0579(5) &0.0580(3)   &0.75 \\
Ti(2)		&0.2640(8) &0.2968(4) &0.2285(8)        &0.2531(9) &0.2131(9) &0.1973(4)  &0.1549(4)  & 0.1557(6) &0.1560(3)   &0.75 \\
Fe(1)		&0.2605 &0.2633 & 0.2608           &0.2469 &0.2395 &0.2142 	   &0.0532 &0.0579 &0.0580 	&0.25 \\
Fe(2)		&0.2640 &0.2968 &0.2285        &0.2531 &0.2131 &0.1973	 &0.1549 &0.1557	&0.1560   &0.25 \\
O(1)		&0.3385(5) &0.2743(6) &0.3203(4)             &0.3252 &0.2894(1) &0.3200(6) 	       &0.0 &0.0 &0.0         &0.50 \\
O(2)		& 0.0086(5) & 0.0086(5) & 0.0086(5)         &-0.0955(6) &-0.0535(9) &-0.0232(8)  &0.2443(1)  &0.2435(3) &0.2386(8)  &1.0  \\
O(3)		&0.3240(2) &0.2891(2) &0.3048(7)         &0.1560(9) &0.2294(3) &0.3519(1)   &0.0757(1) &0.0783(5) &0.0670(4)	&1.0 \\
O(4)		&0.2980(6) &0.3690(8) &0.2799(5)             &0.3012(6) &0.3103(3) &0.3101(6)      &0.1879(9) &0.1974(5) & 0.1918(9)  &1.0 \\
O(5)		&0.0419 &0.0752(6) & 0.0853(4)           &0.0752(1) &0.0099(4) &0.0245(8) 	    &0.0426 &0.0451(7) &0.0520(5)	&1.0  \\
O(6)		&0.6298(8) & 0.7182(7)   & 0.7731(7)       &0.3846(7)	&0.3630(2) &0.3414(2)   &0.0736(8) &0.0749(7) &0.0782(7)	  &1.0 \\
O(7)	  &0.0704(1) & 0.0881(3) & 0.0760(1)        &-0.0317(4) &-0.0133(4) &0.0289(3) &0.1475(5)  &0.1445(1) &0.1505(9)     &1.0 \\
O(8)		& 0.5888(1) & 0.6005(3) &0.5108(3)          &0.5144 &0.5040(7) &0.4346(6)	  &0.1387 &0.1459(5) & 0.1517(1)  &1.0\\

\end{tabular}
\end{ruledtabular}
\label{tab:parameters}
\end{table*}

Pressure dependent Raman spectroscopy data is shown in Figure.2 of the main text. The normalized Raman data is fitted considering Lorentzian line shape (not shown) and thus obtained variation of Raman mode position with pressure is shown in Figure.~\ref{fig:wavenumber}. No anomalies are observed in the variation, however with the principal component analysis (PCA) we could individuate two regions with an evident change around 5.5 GPa which is also in rough agreement with the structural transitions reported by Ferreira et al., as discussed in the main text \cite{Ferreira}. Further, to study the change in vibrational modes with unit cell volume, Gr$\ddot{u}$neisen parameters have been calculated. These parameters describe the elastic behavior of materials with pressure \cite{GP}. Isothermal Gr$\ddot{u}$neisen parameter is calculated from Raman spectroscopy measurement and x-ray diffraction measurement data with pressure in the present work. The lattice parameters observed at high pressure are used to calculate the bulk modulus B$_0$, given by $B_0 = -V_0 \left(\frac{\Delta P}{\Delta V}\right)_T$, where V$_0$ is the initial volume, $\Delta$ V is the change in volume and $\Delta$ P is the change in pressure value. The calculated value of the bulk modulus B$_0$ is found to be around 213.7 GPa. The isothermal Gr$\ddot{u}$neisen parameter are calculated from the equation $\gamma_{iT} = \frac{B_0}{\omega_i} \left(\frac{\partial \omega_i}{\partial P}\right)_T$, where B$_0$ is the bulk modulus, $\omega_i$ is the Raman wave-number frequency and $\left(\frac{\partial \omega_i}{\partial P}\right)_T$ is the variation in vibrational mode frequency with pressure. Thus obtained values of $\gamma_{iT}$ are also mentioned in Figure.~\ref{fig:wavenumber} for the respective Raman modes and indicate that the Raman mode at about 150 cm\textsuperscript{-1} seems to be more sensitive for pressure as it exhibits highest $\gamma_{iT}$ value.

As mentioned in the main text of the manuscript, in-order to look for fine structural details we applied principal component analysis (PCA) to the XRD and Raman spectra independently using Orange Software \cite{Orange} in the present work. The XRD/Raman data have been imported in the 2$\theta$ range of 6-34$^{\circ}$ and 140-950 cm$^{-1}$ range respectively. The background was subtracted (rubber band model) and the data were normalized (vector normalization). In both the XRD and Raman datasets, it is observed that the first 3 PCs account for more than 95\% of the total variance as shown in Figure.~\ref{fig:PCA2}). The XRD/Raman data were partitioned using the K-means \cite{KMeans} algorithm and the silhouette analysis \cite{Silouette} implemented in the Orange program. The silhouette analysis indicated  three statistically significant clusters for the XRD dataset and only two statistically significant clusters for the Raman dataset. The Figure.~\ref{fig:PCA2} shows the PC$_i$ vs PC$_j$ plots (i,j = 1,3) for the XRD and Raman datasets and allows to visually recognize the PC$_{i-j}$ correlations motivating the origin of the clustering. Thus obtained results of PC$_2$ for the XRD and the Raman datasets are shown in Figure.2 of the main text.

\renewcommand{\thefigure}{S2}
\begin{figure}[htbp]
  \centering
  \begin{minipage}[b]{0.49\textwidth}
    \includegraphics[height=6cm, width=\textwidth, keepaspectratio]{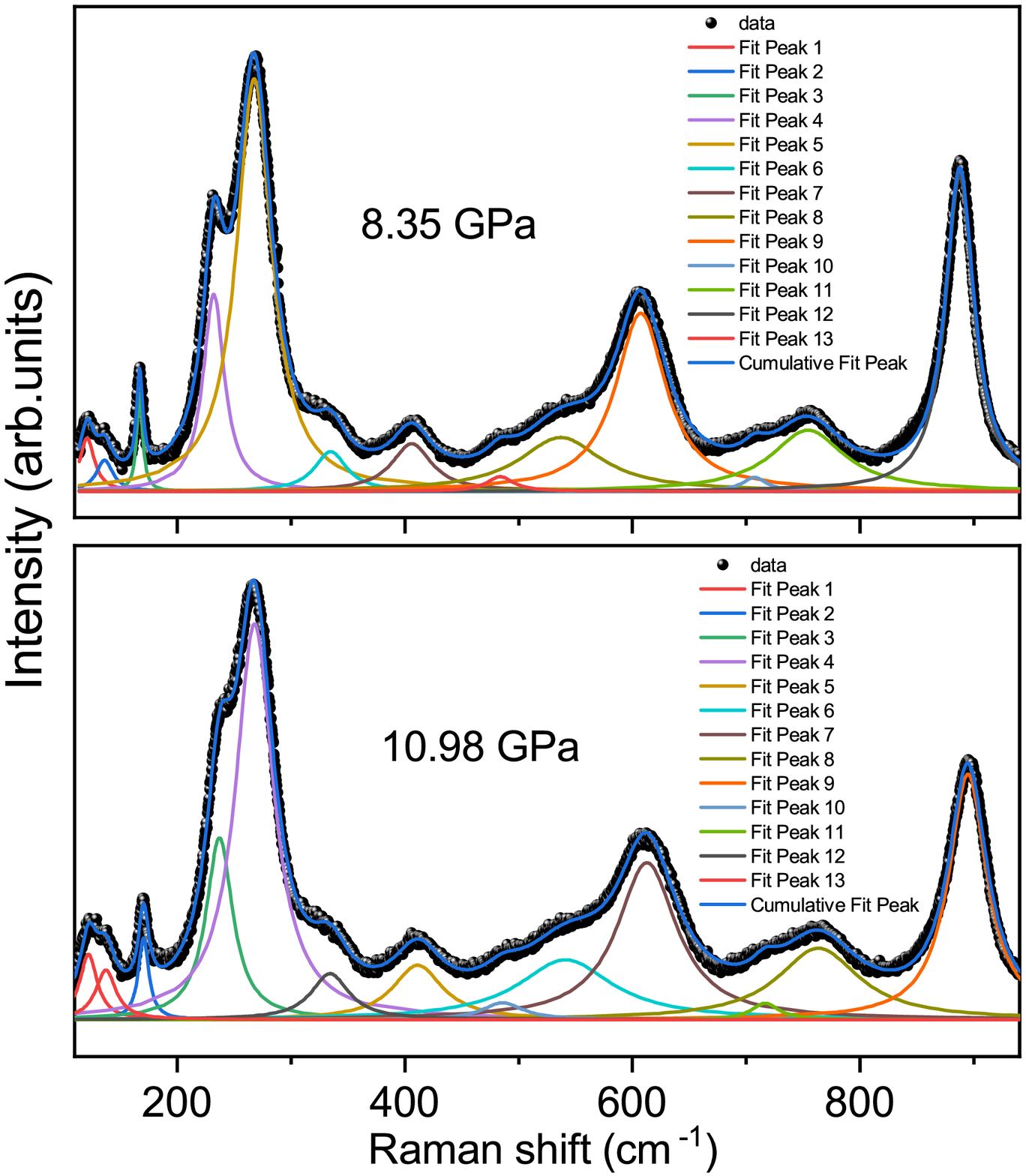}
  \end{minipage}
  \hfill
  \begin{minipage}[b]{0.49\textwidth}
    \includegraphics[height=6cm,width=\textwidth, keepaspectratio]{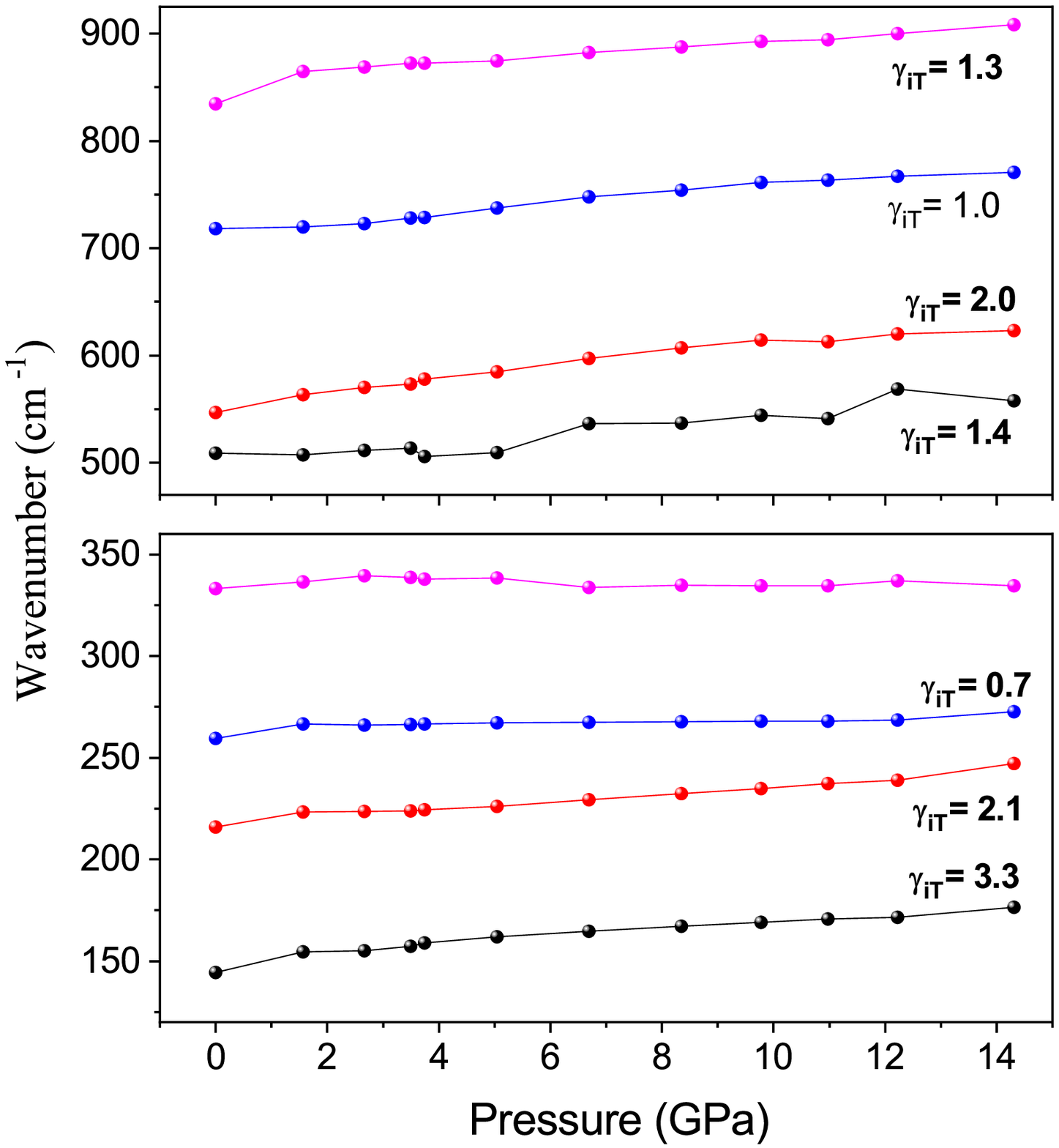}
  \end{minipage}
	\hfill
	\caption{(Left) Representative Lorentzian peak fitted Raman data at 8.35 GPa and 10.98 GPa. Symbol represent the experimental data and the solid lines are the best fit to the data (Right) Representative pressure dependent Raman spectra of polycrystalline Bi\textsubscript{5}FeTi\textsubscript{3}O\textsubscript{15}.  $\gamma_{iT}$ is the Gr$\ddot{u}$neisen parameter.}
	\label{fig:wavenumber}
\end{figure}

\renewcommand{\thefigure}{S3}
\begin{figure}[htbp]
\centering
\includegraphics[height=10cm, keepaspectratio]{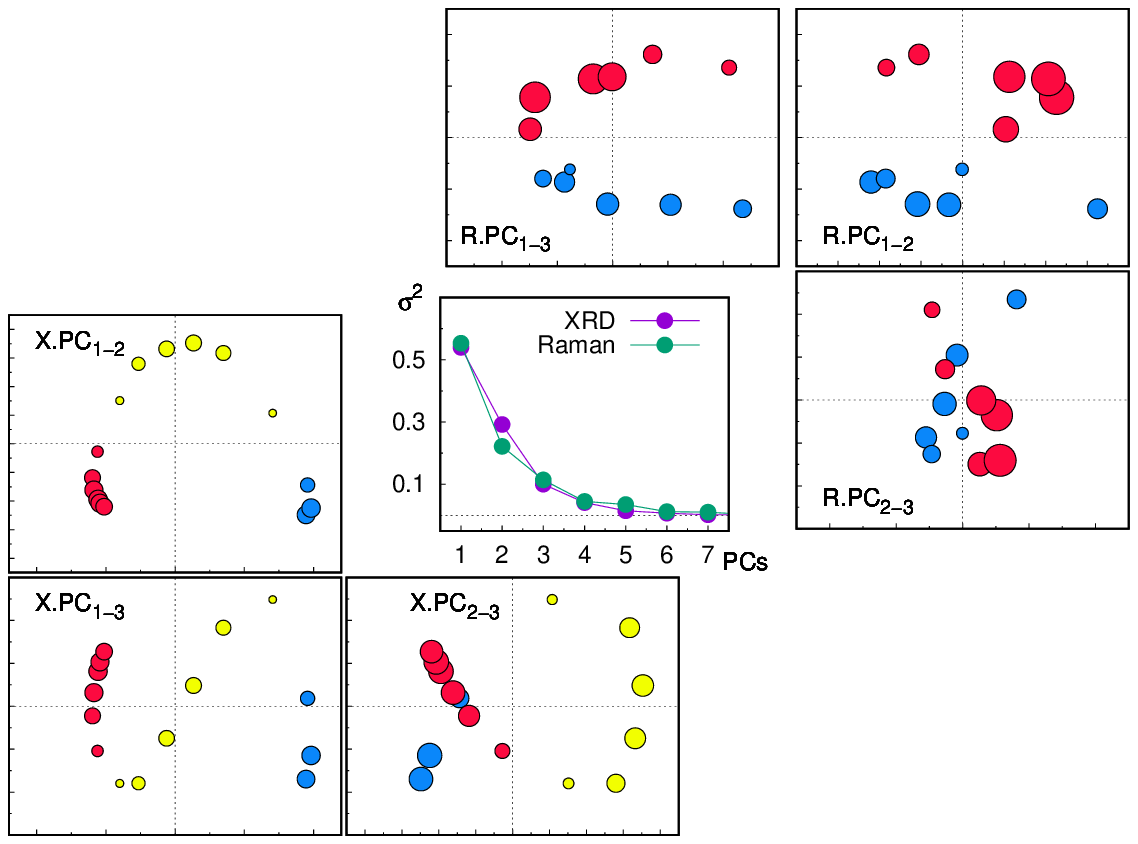}
\caption{The panels show the PC$_i$ vs PC$_j$ plots  for the first three PCs of XRD (X.PC$_{i-j}$) and Raman (R.PC$_{i-j}$) datasets. The colors indicate the clusters individuated by the K-means and the silhouette analysis. The symbol colors specify the cluster assignment while their size is proportional to the silhouette of the point, quantifying the confidence in the assignment of the point to its cluster. In the central panel is reported the scree plots for XRD and Raman datasets which indicates the fraction of variance associated to each PCs.}
\label{fig:PCA2}
\end{figure}


\bibliography{References}